# The Efficacy of a Virtual Reality-Based Mindfulness Intervention

Caglar Yildirim
Khoury College of Computer Sciences
Northeastern University
Boston, MA, US
c.yildirim@northeastern.edu

Tara O'Grady
Department of Computer Sciences
State University of New York at Oswego
Oswego, NY, US
togrady@oswego.edu

*Abstract*—**Mindfulness can be defined as increased awareness of and sustained attentiveness to the present moment. Recently, there has been a growing interest in the applications of mindfulness for empirical research in wellbeing and the use of virtual reality (VR) environments and 3D interfaces as a conduit for mindfulness training. Accordingly, the current experiment investigated whether a brief VR-based mindfulness intervention could induce a greater level of state mindfulness, when compared to an audio-based intervention and control group. Results indicated two mindfulness interventions, VR-based and audio-based, induced a greater state of mindfulness, compared to the control group. Participants in the VR-based mindfulness intervention group reported a greater state of mindfulness than those in the guided audio group, indicating the immersive mindfulness intervention was more robust. Collectively, these results provide empirical support for the efficaciousness of a brief VR-based mindfulness intervention in inducing a robust state of mindfulness in laboratory settings.**

## I. Introduction

Virtual Reality (VR) allows users to view and interact with a computer-generated environment. The technology first emerged over 50 years ago, however, due to high cost and specific hardware capabilities it was not easily accessible to the public [1]. In recent years, this has changed. As the price of headsets decrease and our technologies improve, VR is no longer restricted to expensive laboratories. Several commercially available head-mounted displays (HMDs) have been released, ranging from high-end technologies such as the Oculus Rift and HTC Vive to low-end technologies such as Google cardboard paired with a VR capable smartphone.

With the proliferation of commercial-grade VR headsets, the use of VR in research applications has progressed. VR allows for controlled perceptual stimuli, a controlled environment, and constraint of user interactions. These qualities have great potential to benefit research: extraneous variables that would otherwise be uncontrollable can be limited, making the implementation of VR applicable to a broad range of subjects. For this reason, there has been widespread use and adoption of VR technology in research, including investigations of mindfulness training and meditation [2].

Mindfulness, in its simplest form, entails reflexively attending to and contemplating one's own experience, embodied within the body, feelings, states of mind, and experiential phenomena. Mindfulness has been a topic of empirical research since its introduction in 1979 as a stress reduction intervention by Jon Kabat-Zinn [5]. Since then, there has been an increasing scholarly interest in its applications for empirical research, and mindfulness training has been shown to lead to better cognitive control, increased happiness, satisfaction with life, and overall psychological wellbeing [31].

While previous studies have provided converging evidence for the benefits of long-term mindfulness training interventions, lasting from several weeks to months [5], these studies necessitate the investment of a considerable amount of time, financial resources, and preparation, making it harder to investigate the effects of MT in diverse settings for many researchers. Therefore, it is useful to explore whether brief MT interventions, which involve participants engaging in a single mindfulness practice for a short period of time (5-20 minutes), can yield positive outcomes similar to those obtained from long-term MT interventions [30]. Experimental manipulations aimed at inducing a state of mindfulness do not attempt to replicate the long-term benefits of MT, but instead are being used to examine the short-term effects of brief mindfulness practices on certain dependent variables conceptually linked to mindfulness [32]. Therefore, although such brief interventions may not provide as robust benefits as long-term interventions, these investigations can provide invaluable insight into immediate benefits of brief MT and its application to various domains, including wellbeing, interpersonal relationships, daily life settings, work environments, education, etc. Previous studies into such brief interventions have yielded mixed results [5, 11, 24], underscoring the importance of devising strategies to induce a robust state of mindfulness in controlled experimental settings.

To address the need for a robust method of inducing mindfulness, we explored the potential of VR as an intervention technology to induce mindfulness. We conducted a between-subjects experiment in which participants ($N = 45$) were randomly assigned to one of three conditions: VR-based intervention, audio-based intervention, and active control group. Participants in the VR-based intervention condition engaged in a 10-minuted guided mindfulness practice in a VR environment, while participants in the audio-based intervention condition engaged in the same mindfulness practice, using the audio recording only (without the VR environment). Participants in the active control group listened to an audiobook for 10 minutes. After the practice/audio phase, participants indicated their state mindfulness levels and completed a short attention task. Results indicated that the VR-based intervention induced the greatest level of state mindfulness, when compared to the other two conditions and that there were no immediate beneficial effects of a brief mindfulness practice on mind wandering.

With the current study, our goal was to present a first step in the direction of utilizing VR-based interventions to induce a state of mindfulness and explore its immediate effects in experimental settings. Another aim of the current study was to call for future research into the design, implementation, and evaluation of immersive VR-based interventions to induce and cultivate mindfulness and promote cognitive functioning and overall wellbeing. The contributions of the current study are as follows:

- Implementation of a VR-based mindfulness intervention enriched with such immersive elements as ambient sound effects and guided practice,
- Empirical support for the effectiveness of a VR-based mindfulness intervention in inducing a state of mindfulness, and
- Experimental design strategies and future directions for investigating the efficaciousness of such VR-based interventions.

## II. RELATED WORK

In this section, we review related work examining outcomes of mindfulness interventions and the effects of integrating VR into mindfulness practices.

### A. Mindfulness

Mindfulness can be defined as increased awareness of and sustained attentiveness to the present moment. Theoretical accounts of mindfulness offer an operational definition that consists of two components. The first component emphasizes self-regulated attention; how much focus one has on the immediate experience [4]. The immediate experience can involve several forms of self-monitoring, such as body sensations, emotional reactions, mental images, mental talk, and perceptual experiences [5]. The second component is the curiosity and acceptance of the experience, outside of whether one desires it. Many contemporary approaches to mindfulness stress the importance of embracing an attitude of openness towards one's experience [5].

### B. Benefits of Mindfulness

Mindfulness interventions have been shown to impact physical health, mental health, cognitive and affective states, and even interpersonal relations. Regarding physical health, it has been found that interventions can bolster body awareness, promote relaxation, and improve stress management and coping skills [5]. Studies have demonstrated improved management of chronic pain, as well as reduced symptoms of stress-related conditions such as fibromyalgia, IBS, breast cancer, and psoriasis [5]. For applications to mental health, there is strong evidence that mindfulness interventions reduce depression relapse rates, rumination, anxiety, PTSD symptomology, and improve outcomes in the treatment of drug addiction [5].

The current mindfulness literature has utilized interventions ranging from short-term [11, 12] to long-term interventions [13]. Several studies have shown that using brief interventions can increase mindfulness. Xua et al. [20] compared a guided audio intervention to a control group. They found that ten minutes of brief interventions overall increased mindfulness. Another study conducted by Taraban et al. [3] found that listening mindfully for twelve minutes to sounds of nature increased mindfulness.

Cognitive and affective outcomes of mindfulness interventions reflect that practicing mindfulness can improve working memory and sustained attention [5]. Several studies support the connection between mindfulness and attentional control [4], as evidenced by improvements in the Stroop test, a widely used attention task to assess selective attention and cognitive flexibility [4]. A study performed by Kramer et al. even demonstrated that mindfulness meditation influences the perception of time, highlighting the focus on the present moment [6].

### C. Mind Wandering

Considering mindfulness training involves the self-regulation of attention, mind wandering, the drifting of attention from an ongoing task to task-unrelated thoughts, has been closely linked to mindfulness. Recognizing the potential of MMT as a means of enhancing one's attentional control and executive functioning abilities as well as increasing one's awareness of the present moment, researchers have proposed mindfulness as an antidote for mind wandering and have become interested in its effects on mind wandering [11]. For instance, Mrazek et al. [11] reported two studies investigating the relationship between mindfulness and mind wandering. Study 1 was aimed at disentangling the association between the Mindful Attention Awareness Scale (MAAS), which is a commonly used measure of dispositional mindfulness, and several measures of mind wandering. Results from Study 1 showed that high levels of mindfulness, as measured by the MAAS, were associated with fewer episodes of mind wandering. While Study 1 provided evidence for the notion that mindfulness and mind wandering are opposing constructs, the generalizability of the results was limited by the correlational design of the study. In Study 2, therefore, the authors set out to examine the effects of an induced short-term mindfulness state on mind wandering. Specifically, they were interested in whether a brief (8-minute) mindful breathing exercise could attenuate mind wandering. There were three groups, namely mindful breathing, passive relaxation, and reading. All participants completed the same attention task before and after the manipulation. Compared to the participants in the passive relaxation and reading groups, participants in the mindful breathing group had fewer SART errors and lower reaction time coefficient of variability, which have been associated with reduced mind wandering. Mrazek et al. concluded that a brief 8-minute mindfulness practice could reduce mind wandering during a sustained attention task.

Prior work has demonstrated that similar to long-term mindfulness interventions, short-term interventions can yield promising results regarding the beneficial effects of mindfulness training on attentional control. Mrazek and colleagues found that just eight minutes of breath-focused mindfulness training led to improved performance on a sustained attention task as compared to both reading and passive relaxation control groups [11]. Similarly, Sormaz and colleagues showed that thirty minutes of a guided, breathing mindfulness practice led to considerable enhancements in distractor filtering [12], and Xua et al. [20] found that brief guided audio intervention prevented the increase of mind wandering.

*D. Mindfulness Training in VR*

The integration of technology into mindfulness practices has already demonstrated positive outcomes. Specifically, visual stimuli have been utilized to create an isolated context for relaxation [27]. VR can amplify these positive findings; it allows a user to feel present in a virtual environment (VE) solely designed to enhance the state of mindfulness. The degree of presence is influential in mindfulness research incorporating VR; thus, mindfulness measures are often correlated with a level of presence in the VE.

VR enhances presence through stimuli oriented with how a person perceives their environment. The HMD uses visual input to stimulate the visual cortex, and sound (e.g., sounds of nature, sounds of water, and birds chirping in a VE) to stimulate the auditory cortex. In a recent study by Navarro-Haro et al., participants reported having a moderate to a strong sense of presence while practicing mindfulness skills in a VE. Results demonstrated that after a brief intervention, participants showed a significant increase in mindfulness state as well as an improved emotional state after the VR session [2]. Furthermore, participants were very accepting of the intervention and encouraging the use of VR for mindfulness. These findings are promising given the brief intervention used.

A study focusing on the intersection of meditation and VR by Shaw et al. [14] utilized a VE in which participants experienced three phases. Each phase consisted of a different environment with a "vocal coach" guiding them through different exercises, using skin conductance as biofeedback to alter the environment's state. Results revealed that post-session relaxation levels were significantly higher than pre-session relaxation levels, indicating again that short-term interventions could produce immediate effects [14].

VR may also afford the reduction of mind wandering during mindfulness practice. VR consumes a lot of attentional resources [3], making it less likely that attention will drift to sights and sounds in the real world. The increased sensory input can also reduce distressing thoughts or emotions while in VR, as improved emotional states after brief interventions have been reported [2]. In a study conducted to evaluate a neuroadaptive VR meditation system called *RelaWorld* [15], participants completed six 10-minute virtual meditation sessions. Results indicated that when compared to a computer screen the use of an HMD elicited fewer hindrances, or negative feelings about the meditative experience, a greater sense of relaxation, a higher level of self-reflection, positive transpersonal qualities, and greater feelings of non-duality [15].

*E. The Current Study*

Previous studies have provided support for the potential of VR in increasing relaxation levels. Nonetheless, prior work incorporating VR into mindfulness and/or relaxation practices has not specifically investigated the efficacy of VR-based interventions in inducing a state of mindfulness. Expanding on the potential of VR demonstrated in previous studies and the increased sense of presence afforded by VEs, we formulated the following hypotheses:

- Hypothesis 1a: the two brief mindfulness interventions, VR-based and guided audio, would induce a greater level of state of mindfulness when compared to the control group.
- Hypothesis 1b: the VR-based mindfulness intervention group would state a greater level of mindfulness when compared to the guided audio group.
- Hypothesis 2a: the VR-based mindfulness interventions and the guided audio intervention would induce greater reductions in the level of mind-wandering than for the control group.
- Hypothesis 2b: the reductions in the level of mind-wandering would be greater in the VR-based mindfulness intervention group than in the guided audio group.

III. METHOD

*A. Design*

In the current study, the independent variable was the type of intervention participants received and was composed of three levels: VR-based mindfulness intervention, audio-based mindfulness intervention, and an active control intervention. The independent variable was manipulated between-subjects to minimize demand characteristics. The dependent variables were state mindfulness, as measured by the State Mindfulness Scale [17], and performance markers of mind wandering, as indexed by the Sustained Attention to Response Task [29].

*B. Participants*

The sample was selected from the local undergraduate student population. Forty-eight undergraduate students participated in the experiment in exchange for course credit. Three participants were excluded from the experiment due to program errors during data collection, resulting in a sample of 45 participants (27 females) with a mean age of 20.5 ($SD = 1.57$, age range: 18-26) years old. All participants reported little to no prior experience with mindfulness. This study was reviewed and approved by the local Institutional Review Board prior to the onset of experiments.

*C. Materials*

 *1) Interventions*

In the current experiment, we employed two brief mindfulness interventions: VR-based and audio-based. The two interventions were identical, except for the VR component utilized in the former. The VR-based intervention used a 3D beach environment in which participants could experience Costa Del Sol [28]. The environment featured a sandy beach environment surrounded by rock formations in blue ocean whose waves were visible and populated with palm trees swaying in the breeze (See Fig. 1 for a screenshot of the virtual environment).

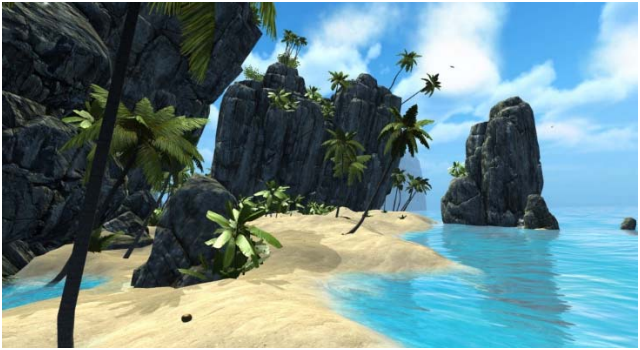

Fig. 1. Screenshot of the Virtual Environment [28]

Participants in the two mindfulness intervention groups, VR-based and audio-based, listened to the same brief guided mindfulness practice designed to prompt the listener to pay attention to his or her breathing and to try to experience the present moment without judgment. This guided mindfulness practice was 10 minutes long. The audio recording included some ambient beach sound effects (ocean wave and bird chirping sounds) to make the practice more immersive. Those assigned to the listening group listened to an audiobook for the same amount of time as the mindfulness training. Specifically, participants in the listening group listened to a portion of the audiobook of JRR Tolkein's *The Hobbit*, which has been used in previous studies [30]. The portion of the audiobook used in the current experiment corresponded to the first chapter of the book and it was 10 minutes long.

 *2) State Mindfulness Scale*

The operational definition of mindfulness accepted by this study consists of two components; emphasis on self-regulated attention and the curiosity and acceptance of the experience. This operational definition is the basis for the development of the State Mindfulness Scale (SMS) used in this study [17]. The SMS is a 21-item measure of state mindfulness and was developed to capture the state-like experience of mindfulness following a mindfulness practice. All participants completed the SMS immediately after the audio recording (guided mindfulness practice for the two mindfulness intervention groups and the audiobook for the active control group), using a 5-point Likert scale. An SMS score was calculated for each participant by calculating the average of the responses to the items on the SMS. Greater SMS scores indicated greater levels of induced mindfulness state.

 *3) Sustained Attention to Response Task*

Participants completed a Sustained Attention to Response Task (SART) following the completion of audio recording and SMS. The SART is a go/no-go sustained attention task in which participants are shown an array of visual stimuli of a single category and are asked to respond to nontargets and to withhold response to targets [29]. In the version of the SART used in the current study, nontargets, or go stimuli, were all the digits except for the digit 3, and targets, or no-go stimuli, were the digit 3. Participants were asked to make a response as fast as possible to all digits except for the digit 3, which were frequently presented, and to withhold response to the digit 3, which was rarely presented. The SART consisted of 225 trials (200 nontargets and 25 targets) and was completed in approximately 4.3 minutes. The presentation time was 250 ms for all stimuli with a mask of 900 ms (i.e., double circles). Fig. 2 illustrates a typical flow of three trials.

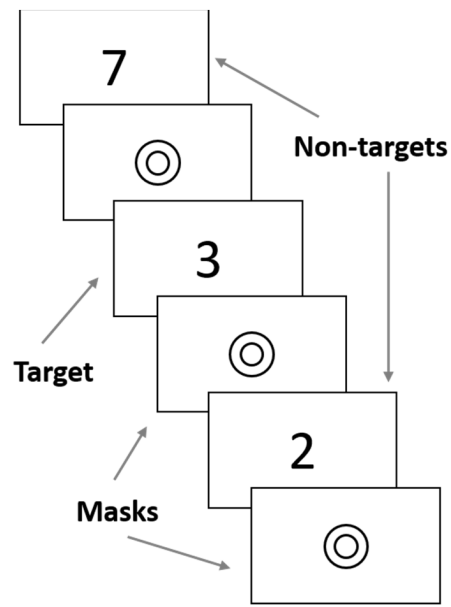

Fig. 2. Example SART trials

 *4) Procedure*

Participants, upon arrival to the lab, provided informed consent, and then were randomly assigned to one of the three intervention conditions: a VR environment with guided audio practice, guided audio practice, or a listening (control) condition. All participants were seated during the entire experiment, regardless of their experimental condition. The VR-based mindfulness practice condition used a commercially available VR headset, HTC Vive. Using the Guided Meditation VR application on Steam [19], participants were virtually teleported to a beach environment that provided participants a visual experience with sand, lapping water against a coastline, and palm trees swaying in the wind. The audio provided for this condition was a guided mindfulness practice that instructed participants to focus on their breathing. The guided audio lasted for ten minutes. Similar to the VR group, the guided audio condition used the same guided mindfulness practice audio (but without the VR environment). The listening (control) group listened to the first chapter of JRR Tolkien's *The Hobbit*, which also lasted for ten minutes. All conditions used the same headphones for audio output. After the audio phase, participants completed the SMS to report their level of state mindfulness. Then they were informed about the SART. Participants were asked to press the space key bar as quickly as possible when the digit was not 3 and to refrain from responding when the digit was 3. Having completed some practice trials, participants began completing the SART. At the end of the experiment, participants were debriefed about the purpose of the study.

## IV. RESULTS

We conducted several descriptive and inferential statistical tests to examine the hypotheses stated earlier. Table 1 presents a summary of these tests for each of the dependent variables.

TABLE I. SUMMARY OF HYPOTHESIS TESTING

|  | $M$ ($SE$) | $F$ | $MSE$ |
|---|---|---|---|
| **State Mindfulness** |  | 8.65* | .596 |
| VR-Based | 3.54 (.186) |  |  |
| Guided Audio | 3.01 (.258) |  |  |
| Listening | 2.37 (.133) |  |  |
| **SART Errors** |  | .559 | .043 |
| VR-Based | .425 (.055) |  |  |
| Guided Audio | .478 (.054) |  |  |
| Listening | .399 (.055) |  |  |
| **RT CV** |  | .169 | .053 |
| VR-Based | .378 (.064) |  |  |
| Guided Audio | .354 (.060) |  |  |
| Listening | .327 (.060) |  |  |

* $p < .001$. $n = 15$ in each condition (total $N = 45$).
State Mindfulness represents the average SMS score. SART Errors represent the percentage of errors in withholding response. RT CV represents the variation in reaction times.

Hypothesis 1a was that the two brief mindfulness interventions would induce a greater state of mindfulness as measured by the average score on SMS, when compared to the listening group. Hypothesis 1b was that of the two mindfulness interventions, VR-based intervention would lead to a greater state of mindfulness (see Fig. 3). A one-way analysis of variance (ANOVA) was conducted to examine the effect of experimental manipulation on SMS scores. The assumption of homogeneity (equality) of variances, as assessed by Levene's test ($p > .05$) was met. Results revealed a statistically significant difference among the three conditions, $F(2, 42) = 8.65$, $MSE = .596$, $p < .001$, $\eta^2 = .29$, $\omega^2 = .25$, $BF_{10} = 46.17$. Considering the effect size measures, the amount of variance in SMS scores explained by the manipulation of mindfulness intervention was substantially large. In line with Hypothesis 1a, planned contrasts indicated that the two mindfulness interventions, VR-based and guided audio, induced a greater state of mindfulness, $t(42) = 3.71$, $p < .001$ (one-tailed), when compared to the control group. Similarly, in line with Hypothesis 1b, planned contrasts revealed that participants in the VR-based mindfulness intervention group reported a greater state of mindfulness than those in the guided audio group, $t(42) = 1.88$, $p < .05$ (one-tailed).

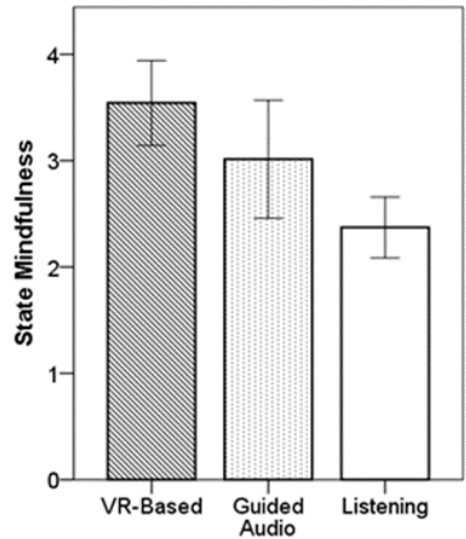

Fig. 3. Average state mindfulness levels as measured by the SMS on 5-point Likert Scale. Greater scores indicate greater levels of state mindfulness. Error bars represent 95% confidence intervals.

Hypotheses 2a and 2b were concerned with the changes in behavioral markers of mind wandering before and after the intervention. More specifically, Hypothesis 2a predicted greater reductions in SART errors and reaction time coefficient of variability, RT CV, for the two mindfulness groups than for the control group. Hypothesis 2b was that these reductions would be greater in the VR-based mindfulness intervention group than in the guided audio group. To test these hypotheses, a one-way analysis of covariance (ANCOVA) was conducted, controlling for the proportion of SART errors measured at the baseline (before the participants were exposed to the experimental manipulation). Results revealed no significant differences in the proportion of SART errors among the three conditions after adjustment for baseline proportion of SART errors, $F(2, 41) = .559$, $MSE = .043$, $p = .576$, $\eta^2_p = .03$, $\omega^2 = .00$, $BF_{01} = 3.29$. Similarly, ANCOVA results revealed no significant differences in RT CV among the three conditions after adjustment for baseline RT CV, $F(2, 41) = .169$, $MSE = .053$, $p = .845$, $\eta^2_p = .01$, $\omega^2 = .00$, $BF_{01} = 5.66$. These results suggest that the brief mindfulness interventions produced no immediate improvements in sustained attention performance and thus no reductions in mind wandering, contradictory to predictions.

## V. Discussion

The purpose of the current study was to investigate the efficacy of a VR-based mindfulness intervention in inducing a state of mindfulness and in reducing mind wandering. Results indicated that both VR-based and audio-based interventions induced a greater level of state mindfulness, when compared to an active control intervention (i.e., listening to an audiobook). In both interventions, the same guided-audio recording was used to encourage participants to pay attention to their breathing, thoughts, feelings, and sensations during the practice. It could be argued that engaging in such a contemplative practice is sufficient for inducing a state of mindfulness, with no need for a VR component. Results, however, showed that participants undergoing the VR-based intervention reported greater levels of state mindfulness than did participants in the audio-based intervention, lending support for the contributions of a VR-based immersive practice above and beyond an audio-based guided practice. In line with Hypothesis 1a and 1b, these results indicate that a brief mindfulness practice is effective in inducing a state of mindfulness in laboratory settings and that VR-based interventions lead to greater levels of state mindfulness.

Another portion of our study focused on mind wandering through attention tasks, where results showed no significant differences in two performance markers of mind wandering, SART errors and RT CV, among the three conditions. Our hypothesis predicted greater reductions in performance markers of mind wandering for the VR and audio conditions than for the control group. However, this was inconsistent with the results. The hypothesis further implied that these reductions would be greater in the VR-based mindfulness intervention group than in the guided audio group, which was also not supported by the results. This suggests that the brief mindfulness interventions produced no immediate improvements in sustained attention performance and thus no reductions in mind wandering, contrary to our predictions.

The inconsistencies with previous findings that mindfulness interventions improve sustained attention could be attributed to several factors. Studies that have demonstrated improvement in sustained attention involve mindfulness interventions that differ from the guided mindfulness intervention used in our study. Mrazek et al. [11] demonstrated that eight minutes of mindful breathing led to reduced behavioral indicators of mind wandering. Zeidan et al. found that after four sessions of mindfulness intervention delivered over a four-day span improved performance on tasks related to sustained attention [22]. Both of these studies implemented a mindful breathing task. Guidance was provided in parts of the study carried out by Zeidan et al. [22]. However, the majority of the time was spent in silence meditating. Additionally, longitudinal research that involved guidance and training of mindful breathing practices over the course of three months reported significant improvement in sustained attention [23]. While participants in our study were able to attain a significant state of mindfulness, it is possible that guided meditation in both the virtual and audio conditions prevented participants from attaining the necessary reduction in mind-wandering to improve performance on the sustained attention task within the short time frame.

The vast majority of studies conducted on mindfulness are field studies with clinical populations, such as individuals with eating, mood, anxiety, or personality disorders [9]. Although these studies demonstrate the benefits of mindfulness-based practice, there is a lot to be learned about the underlying effects of mindfulness on well-being [9]. There is evidence that mindfulness training causes neurobiological changes [10]. However, the validity of findings at neurological levels such as changes in alpha and theta brain waves and cortical thickening are debated; this is due to the low number of clinical studies that have been performed, the small number of participants in these studies, and the lack of longitudinal design [7]. Nonetheless, studies have found that participants who received a brief mindfulness intervention prior to a verbal learning test recollected a significantly lower proportion of negative words than positive when compared to a control group on a delayed recall task [9]. These findings demonstrate that beyond the benefits individuals can gain from mindful practices, mindfulness interventions can provide an avenue to gain insight into a further understanding of cognitive functions.

### A. Future Research

The current study indicates that further investigations into the use and efficacy of VR-based mindfulness interventions and the influence of these interventions on mind wandering are warranted. For instance, future research could investigate the influence of a more thorough, but still brief, mindfulness intervention on mind wandering. Such an investigation could have participants engage in a brief mindfulness practice and then discuss with the participants what they experienced during the practice. Having ensured that they understand the importance of focusing on their breathing and of nonjudgmentally bringing attention back to their breathing whenever their minds wander off, the experimenters could have the participants engage in another brief mindfulness practice, after which the participants could go on to the attention task. In so doing, the empirical question is whether this type of modified brief intervention might reduce mind wandering.

Given the immediate beneficial effects of the brief intervention on state mindfulness, it would be especially useful for future studies to focus on short-term mindfulness interventions (e.g., multiple practices over several days or a few weeks) and to incorporate a self-regulated training regimen that allows participants to practice mindfulness repeatedly and consistently over an extended period of time, as opposed to a one-time practice. Using such a strategy, it would be possible to investigate the effects of VR-based mindfulness interventions on both state and trait mindfulness, mind wandering, and psychological wellbeing. Cahn and Polich, for instance, provided support that consistent and extensive meditation training promotes lasting changes in cognitive abilities and well-being [26].

One limitation of the current study concerns the extent to which individuals in the mindfulness meditation group could immerse in the brief meditation practice. A limitation inherently present in most contemplative studies, it is not possible to behaviorally determine the extent to which individuals are able to focus on the present moment during the practice without using physiological measures. This limitation may be even more pronounced for brief mindfulness interventions, in which individuals have little guidance as to how well they are engaging in the meditation practice. Therefore, future studies could address this limitation by incorporating physiological measures into the experimental procedures to assess individuals' engagement level with the mindfulness practice. Such measures can also be useful for the purposes of providing real-time feedback to the individuals regarding their current state of being and helping them better focus on the chosen object of attention or the anchor (i.e., one's breathing).

In addition, future research could consider a comparison between novice and experienced mindfulness practitioners and investigate the efficacy of a VR-based intervention in inducing a state of mindfulness. It would be interesting to test whether the VR-based intervention, being a more immersive practice, would induce a state of mindfulness in novice practitioners similar to the level of state mindfulness reported by experienced practitioners.

Another line of future research would be to develop and validate an adaptive mindfulness intervention in VR (e.g., [33]), where the virtual environment and guided practice can respond to users' physiological state (breathing or attention patterns). Such an intervention could run users' physiological data through a machine learning pipeline online (as the users are engaged in the practice) and could change how the virtual environment responds, providing virtual biofeedback to the users. This approach could increase users' immersion in the practice and could potentially yield better gains, which is a hypothesis that could be tested in future studies.

*B. Experimental Design Strategies*

One aspect of the current study that differed from similar studies on the incorporation of VR into mindfulness and relaxation practices [2, 14, 15, 19] is the use of two treatment conditions and one active control conditions. More specifically, instead of comparing the VR-based mindfulness intervention to a control group, the current experiment incorporated both VR-based and audio-based interventions, which enabled us to rule out the possibility that engaging in a mindfulness practice was sufficient for inducing a state of mindfulness. Additionally, the current study utilized an active control group in which participants listened to an audiobook for the same amount of time the mindfulness interventions lasted, 10 minutes. By incorporating such an active control group, the current experiment concluded that listening to an audio recording was not enough for inducing a state of mindfulness. Instead, the current experiment showed that a state of mindfulness was achieved only after engaging in a mindfulness practice and that greater levels of state mindfulness could be obtained by completing the practice in VR than by completing the practice using guided-audio recordings without an immersive virtual environment. Future studies could take these suggestions into consideration when planning and designing similar intervention studies.